\documentclass[paper,prd,reprint,preprintnumbers,nofootinbib,notitlepage,showpacs,showkeys]{revtex4-1}
\usepackage{latexsym,graphicx,amssymb,amsmath,mathrsfs} 

\DeclareSymbolFont{bbold}{U}{bbold}{m}{n}
\DeclareSymbolFontAlphabet{\mathbbold}{bbold}

\newcommand{\be}{\begin{equation}}      
\newcommand{\ee}{\end{equation}}      
\newcommand{\bea}{\begin{eqnarray}}      
\newcommand{\eea}{\end{eqnarray}}    
     
\newcommand{\rt}[1]{{}}      
\newcommand{\crit}{\,\textrm{crit}\,}

\newcommand{\Tr}{\,\textrm{Tr}\,}

\newcommand{\gf}{\,\textrm{gf}\,}

\newcommand{\gh}{\,\textrm{gh}\,} 
\newcommand{\rhs}{\,\textrm{rhs}\,}  
\newcommand{\lhs}{\,\textrm{lhs}\,}  
\newcommand{\mult}{\,\textrm{multiplicity}\,} 
 
\newcommand{\const}{\,\textrm{const.}\,}

\newcommand{\Rone}{\,\textrm{R1}\,} 
\newcommand{\Rtwo}{\,\textrm{R2}\,}

\makeatletter
\renewcommand\appendix{\par
\setcounter{section}{0}%
\setcounter{subsection}{0}%
\gdef\thesection{\appendixname\space\@Alph\c@section}}

\long\def\unmarkedfootnote#1{{\long\def\@makefntext##1{##1}\footnotetext{#1}}}
\makeatother

\begin{document} 

\title{Renormalization group flows of the $N$-component Abelian Higgs model} 
\author{G. Fej\H{o}s}
\email{fejos@rcnp.osaka-u.ac.jp}
\affiliation{Research Center for Nuclear Physics, Osaka University, Ibaraki, Osaka 567-0047, Japan}
\affiliation{Theoretical Research Division, Nishina Center, RIKEN, Wako 351-0198, Japan}
\author{T. Hatsuda}
\email{thatsuda@riken.jp}
\affiliation{iTHES Research Group and iTHEMS Program, RIKEN, Wako 351-0198, Japan}
\affiliation{Theoretical Research Division, Nishina Center, RIKEN, Wako 351-0198, Japan}

\begin{abstract}
{Flows of the couplings of a theory of an $N$-component (complex) scalar field coupled to electrodynamics are investigated using the functional renormalization group formalism in $d$ dimensions in covariant gauges. We find charged fixed points for any number of components in $d=3$, in accordance with the findings of [G. Fejos and T. Hatsuda, Phys. Rev. D {\bf 93}, 121701 (2016)] for $N=1$. It is argued that the appropriate choice of the regulator matrix is indispensable to obtain such a result. Ward-Takahashi identities are analyzed in the presence of the regulator, and their compatibility with the flow equation is investigated in detail.}
\end{abstract}

\maketitle

\preprint{RIKEN-QHP-312}

\section{Introduction}

Analyses of the phase transition in scalar electrodynamics (also known as the Abelian Higgs model) have a long history. It first became of interest via the Coleman-Weinberg mechanism, which is now a textbook example of how radiative corrections can generate spontaneous symmetry breaking in 3+1 dimensions \cite{rivers}. The same mechanism lead to the conclusion that in the dimensionally reduced theory, which describes a superconductor close to its critical point, only first order transitions may occur \cite{halperin74}. The original argument was backed by renormalization group (RG) analyses using the $\epsilon$ expansion, which showed no infrared (IR) stable fixed point and concluded the absence of critical behavior in the system. When one considered the $N$-component extension of the scalar field, one found traces of charged fixed points (i.e., those with nonzero gauge coupling), but the necessary condition in $d=3$ turned out to be $N\geq 183$ \cite{zinnjustin}, which certainly did not apply for superconductors.

Later on, based on a duality argument, it was shown that the superconducting phase transition might be of second order after all \cite{kleinert82}. A disordered field theory, dual to the original Abelian Higgs model, predicted that the system is capable of showing critical behavior, given that the ratio of the gauge and scalar couplings is small enough. This result was also confirmed by separate Monte Carlo simulations \cite{bartholomew83,kajantie98,mo02}, and the transition was conjectured to belong to the XY universality class \cite{olsson98,kleinert06}. The two component case was also studied in \cite{kuklov08,herland13}, and the question of the order of the transition in Ginzburg-Landau-type models has also drawn attention in systems with various symmetries and interactions \cite{meier15,sellin16,galteland16}.

Going back to the conventional superconductor, the order of the transition, however, could not be understood properly from a RG point of view. In a series of papers there were promising attempts to find IR stable charged fixed points that could describe the second order nature of the superconducting transition \cite{kolnberger90,reuter94,bergerhoff96,bergerhoff96b,folk96,herbut96,freire01,kleinert03}, but it is believed that the fixed point structure is not well understood quantitatively. 

In this paper we present a study of obtaining the flows of the couplings in the $N$-component Abelian Higgs model via the functional renormalization group (FRG) approach in arbitrary dimensions $d$, based on our earlier results \cite{fejos16} for $N=1$. FRG is a generalization of the Wilsonian RG in the sense that, in general, not only flows of individual couplings are considered, but also that of the complete quantum effective action itself \cite{wetterich93,kopietz}. The contents to be presented here are most closely related to \cite{bergerhoff96,bergerhoff96b}, with several differences featured. The most important one is that these earlier works tackled the problem of gauge symmetry violation (due to the momentum cutoff) via the background field formalism, while we are presenting a method, where only one gauge field is included, and gauge symmetry violation is taken care by appropriate gauge fixing and consistency conditions. (Recently this direction is also under investigation in, e.g., \cite{wetterich16}.) Furthermore, as opposed to the earlier studies, we do not rely on any numerics at all, our results are completely analytic. A related difference lies in the inclusion of different variants of Litim's IR regulator \cite{litim01}, as opposed to smooth ones employed in \cite{bergerhoff96,bergerhoff96b}. This turns out to be a crucial point in our reasoning about the existence of charged fixed points. A main advantage of the FRG framework is that one is free to choose between IR regularization schemes, which opens up the possibility of optimizing the RG flows leading to more reliable approximate solutions. We will argue that such an investigation is required for finding the same structure of tricritical and IR fixed points for any value of $N$ in $d=3$.

The question of the Ward-Takahashi identities (WTIs) and gauge symmetry violation is also of particular importance. Approximate solutions of the RG flows should not violate the WTIs to be physically trustworthy. It is well known that in FRG the WTIs are unavoidably spoiled by the regulator terms, but ultimately should recover when fluctuations are integrated out. The problem is that even though any WTI modified by the regulator is compatible with the flow equation of the effective action in the sense that if they are satisfied at a given scale then they are satisfied at all scales \cite{litim98,gies12}, truncations may ruin the virtue of the construction and one has to solve the flow equation and the WTIs simultaneously by dividing operators into WTI dependent and independent classes \cite{ellwanger96,gies04,gies12}. It turns out that in the present case the situation is not that complicated, and the modified WTIs are compatible with our approximation of the flow equation.

The structure of the paper is as follows. In Sec. II, we introduce the model, the gauge fixing, and the truncation scheme. We also establish the general framework of the modified Ward-Takahashi identities (mWTI). Sec. III consists of three parts: first we calculate the flow equations of couplings and wave function renormalizations, and discuss how to accommodate the flow equation and the used ansatz for the effective action by choosing appropriate gauge fixing. Second, we investigate and draw attention on the appropriate choice of the regulator matrix, and third, we present the fixed point analysis itself. Sec. IV is devoted for evaluating the mWTI in the background of classical gauge and scalar fields, where we show that besides boundary conditions, the flow equation and the mWTI contain the same information on the flows of the couplings. The reader finds conclusions in Sec. V.

\section{Model and method}

\subsection{$N$-component Abelian Higgs model}

The model we are investigating is a theory of an $N$-component complex scalar field equipped with $U(1)$ gauge symmetry. In $d$-dimensional Euclidean space the classical action takes the following form:
\bea
\label{Eq:clS}
S=\int_x {\cal L}&=&\int d^dx \Big[\frac14 F_{ij} F_{ij}+(D_i \phi^a)^\dagger D_i \phi^a \nonumber\\
&&+m^2\phi^{\dagger a}\phi^a+\frac{\lambda}{6}(\phi^{\dagger a} \phi^a)^2 \Big],
\eea
where $D_i=\partial_i - ieA_i$ is the covariant derivative with $A_i$ being the gauge field ($i=1,...d)$, $F_{ij}=\partial_i A_j-\partial_j A_i$ is the $U(1)$ field strength tensor, $\phi^a$ is an $N$-component complex scalar ($a=1,...N$), and $m^2$ and $\lambda>0$ are constants. The Lagrangian is invariant under the following gauge transformation:
\bea
\label{Eq:trf}
\delta \phi^a(x) = ie\theta(x)\phi^a(x), \quad \delta A_i(x) = - \partial_i \theta(x),
\eea
where $\theta(x)$ is an infinitesimal spacetime dependent parameter. It is convenient to separate the $\phi^a$ fields as $\phi^a=(\sigma^a+i\pi^a)/\sqrt{2}$, where $\sigma^a$ and $\pi^a$ are real.
We employ the $R_\xi$ gauge fixing condition by adding the following term to ${\cal L}$:
\bea
\label{Eq:gf}
{\cal L}_{\gf}=\frac{1}{2\xi}(\partial_i A_i+\xi e\tilde{\sigma}^a\pi^a)^2,
\eea
where $\xi$ is the gauge fixing parameter, and $\tilde{\sigma}^a$ is a freely adjustable field. It will be useful to tune $\tilde{\sigma}^a$ to the classical field corresponding to $\sigma^a$. The derivative of the gauge fixing condition (\ref{Eq:gf}) with respect to the gauge transformation parameter is field dependent; therefore, ghost fields ($c^*$, $c$) have to be introduced. The corresponding dynamics is described by
\bea
\label{Eq:gh}
{\cal L}_{\gh}=c^*(-\partial^2+\xi e^2 \tilde{\sigma}^a \sigma^a)c.
\eea
In the FRG formalism one derives an evolution equation for the scale-dependent effective action $\Gamma_k$, which incorporates fluctuations with momenta higher than the course graining scale $k$. Following the procedure of the standard perturbative renormalization, we rescale the fields and the charge as follows:
\bea
\phi^a \rightarrow Z_{\phi,k}^{1/2}\phi^a, \quad A_i \rightarrow Z_{A,k}^{1/2} A_i, \quad e \rightarrow \frac{Z_{e,k}}{Z_{A,k}^{1/2}Z_{\phi,k}}e,
\eea
where the $Z_{i,k}$ $(i=\phi,A,e)$ are scale-dependent rescaling factors. The scale-dependent effective action is then approximated with the form of (\ref{Eq:clS}) [and by adding (\ref{Eq:gf}) and (\ref{Eq:gh})], but with scale-dependent couplings and field rescalings:
\bea
\label{Eq:gammaansatz0}
\Gamma_k=\int_x {\cal L}_k&=&\int d^dx\bigg[\frac{Z_{A,k}}{4}F_{ij}F_{ij}+Z_{\phi,k}(\hat{D}_i\phi^a)^\dagger \hat{D}_i\phi^a\nonumber\\
&+&\frac{Z_{\phi,k}m_k^2}{2}\phi^{\dagger a}\phi^a+\frac{Z_{\phi,k}^2\lambda_k}{6}(\phi^{\dagger a}\phi^a)^2 \nonumber\\
&+&\frac{Z_{A,k}}{2\xi_k}(\partial_i A_i+\xi_k \frac{Z_{e,k}}{Z_{A,k}}e\tilde{\sigma}^a\pi^a)^2 \nonumber\\
&+&c^*(-\partial^2+\xi_k \frac{Z_{e,k}^2}{Z_{A,k}Z_{\phi,k}} e^2 \tilde{\sigma}^a \sigma^a)c\bigg],
\eea
where $\hat{D}_i=\partial_i-i\frac{Z_{e,k}}{Z_{\phi,k}}eA_i$, and we have also allowed the gauge fixing parameter ($\xi_k$) to flow. In order to preserve gauge invariance, one needs $\hat{D}_i\phi^a=D_i\phi^a$, which can be achieved by $Z_{e,k}=Z_{\phi,k}$. This equality is indeed a consequence of the Ward-Takahashi identities, but in the FRG scheme it represents a subtle issue, and we will address the question in Sec. IV. At this point we only assume that $Z_{e,k}=Z_{\phi,k}\equiv Z_k$. Now, expression (\ref{Eq:gammaansatz0}) in terms of the collection of fields $\Phi=(A_i,\sigma^a,\pi^a,c^*,c)$ is the following:
\bea
\label{Eq:gammaansatz}
\Gamma_k[{\Phi}]&=&\int_x {\cal L}_k=\int_x\bigg[ \frac{Z_{A,k}}{2}{A}_i[-\partial^2\delta_{ij}+\partial_i\partial_j (1-\xi_k^{-1})]{A}_j \nonumber\\
&+&\frac{Z_{k}}{2}{\sigma}^a(-\partial^2+m_k^2){\sigma}^a+\frac{\lambda_kZ_{k}^2}{4!}\Big(({\sigma}^a)^2+({\pi}^a)^2\Big)^2\nonumber\\
&+&\frac{Z_{k}}{2}{\pi}^a\Big((-\partial^2+m_k^2)\delta^{ab}+\frac{\xi_k Z_{k}}{Z_{A,k}}e^2\tilde{\sigma}^a\tilde{\sigma}^b\Big){\pi}^b \nonumber\\
&+&\frac12 Z_{k}e^2{A}_i{A}_i\Big(({\sigma}^a)^2+({\pi}^a)^2\Big) \nonumber\\
&-&Z_{k}e\partial_i {A}_i ({\sigma}^a -\tilde{{\sigma}}^a)\pi^a-2Z_{k}e{\pi}^aA_i\partial_i {\sigma}^a\nonumber\\
&+&{c}^*\Big(-\partial^2+\xi_k\frac{Z_{k}}{Z_{A,k}}e^2{\tilde{\sigma}}^a {\sigma}^a\Big){c}\bigg].
\eea
The scale-dependent $\Gamma_k$ effective action obeys the following RG-evoution equation \cite{kopietz}:
\bea
\label{Eq:floweq}
\partial_k \Gamma_k = \frac12 \int \Tr [(\Gamma_{k,2}+{\cal R}_k)^{-1} \partial_k {\cal R}_k],
\eea
where $\Gamma_{k,2}=\delta^2 \Gamma_k/\delta \Phi^\dagger \delta \Phi$ is the second derivative matrix, and ${\cal R}_k$ is the so-called regulator, which appears by adding a regulating term,
\bea
\int \Phi^\dagger {\cal R}_k \Phi
\eea
to the action (\ref{Eq:clS}), and which plays the role of suppressing low momentum fluctuations. The dimension of the matrices is $d+2N+2$. Note that, one is allowed to switch basis between $({\sigma}^a,{\pi}^a) \leftrightarrow ({\phi}^{\dagger a}, {\phi}^a)$ for conventional purposes. It is common to rewrite (\ref{Eq:floweq}) as
\bea
\label{Eq:floweq2}
\partial_k \Gamma_k = \frac12 \tilde{\partial}_k \int \Tr \log (\Gamma_{k,2}+{\cal R}_k),
\eea
where $\tilde{\partial}_k$ acts only on ${\cal R}_k$. The structure of this form indicates that the change of the effective action with respect to the coarse graining scale $k$ is given by one-loop diagrams made up by dressed propagators with the inclusion of the regulator.

The evaluation of the flow equation (\ref{Eq:floweq2}) is done via the spectral decomposition of the $(\Gamma_{k,2}+{\cal R}_k)$ matrix. In terms of eigenvalues $\gamma_{k,2}^{(i)}$ of $\Gamma_{k,2}$, assuming a quasilocal expansion of $\Gamma_k$ (i.e., $\Gamma_k$ is considered as an integral of a local density function that depends on fields and their derivatives), we can always rewrite (\ref{Eq:floweq2}) as (evaluating the eigenvalues explicitly in Fourier space)
\bea
\label{Eq:flowlog}
\!\!\!\!\partial_k \Gamma_k = \frac12 \tilde{\partial}_k \int_x \int_{q}\sum_{i=1}^{2N+d+2} \log [\gamma_{k,2}^{(i)}(q) + R^{(i)}_k(q)],
\eea
where we have chosen the ${\cal R}_k$ matrix such that it diagonalizes together with $\Gamma_{k,2}$, and the corresponding eigenvalues are $R_k^{(i)}$. Note that in the rhs of (11), contributions of Grassmannian fields have to be taken into account with an additional minus sign.

We are interested in the flows of the couplings and wave function renormalizations. To extract the corresponding information we have to evaluate the right-hand side of (\ref{Eq:flowlog}) in an appropriate background of a classical field, and identify the operators whose $k$ derivative appears in the left-hand side via the ansatz (\ref{Eq:gammaansatz}). The identifiable operators and the corresponding classical fields that are need to be imposed are the following:
\begin{subequations}
\bea
\label{Eq:bg1}
\frac{Z_k m_k^2}{2}{\sigma}^a{\sigma}^a, \frac{Z_k^2 \lambda_k}{4!} ({\sigma}^a {\sigma}^a)^2\quad \longleftrightarrow \quad
\begin{cases}
{\sigma}^a = \const \neq 0 \\ {\pi}^a=0 \\ {A}_i=0 \\
\end{cases}
\eea
\bea
\label{Eq:bg2}
\frac{Z_k}{2}\sigma^a (-\partial^2) \sigma^a  \quad \longleftrightarrow \quad
\begin{cases}
\partial_j \sigma^a \neq 0 \\ {\pi}^a=0 \\ {A}_i=0 \\
\end{cases}
\eea
\bea
\label{Eq:bg3}
\frac{Z_{A,k}}{2}A_i(-\partial^2\delta_{ij}+\partial_i\partial_j)A_j \quad \longleftrightarrow \quad
\begin{cases}
{\sigma}^a =0 \\ {\pi}^a=0 \\ \partial_j A_i\neq 0.  \\
\end{cases}
\eea
\end{subequations}
Using these identifications first one obtains the flows of combinations $Z_km_k^2$ and $Z_k^2\lambda_k$, then the flow of $Z_k$ itself. This leads to the flows of $m_k^2$, $\lambda_k$, and finally one gets the flow of $Z_{A,k}$ directly. Since the flowing charge can be defined as $e_k^2=e^2/Z_{A,k}$, once the former quantities are given, one is able to search for fixed points.

\subsection{Modified Ward-Takahashi identities}

Before we start to evaluate the flow equation in the background of various classical fields, it is worth establishing the general framework of the (modified) Ward-Takahashi identities in the Abelian Higgs model.

Ward-Takahashi identities play an important role in quantum electrodynamics as gauge invariance is encoded in them after losing explicit gauge symmetry due to the gauge fixing term. One may emphasize the identity between matter field and charge-rescalings ($Z_{\phi,k}=Z_{e,k}$) and the fact that only the transverse component of the gauge propagator receives radiative corrections. In the FRG formalism, this is all lost due to the regulator term, which breaks gauge invariance explicitly. Nevertheless, one can derive mWTIs including additional contributions arising from the regulator, which have to function as consistency relations at intermediate scales, and become the regular Ward-Takahashi identities in the $k\rightarrow 0$ limit.

It is easy to show that the flow equation and the mWTI are compatible with each other in the sense that if at a certain scale the system is on a trajectory where the mWTI is satisfied, then it will remain true throughout the flow. It is argued, however, that introducing truncations may ruin the aforementioned virtue of the flow equation and the standard recipe is to introduce independent operators evolving via the flow equation and use mWTI for dependent ones \cite{ellwanger96,gies12,gies04}.

The standard derivation of the Ward-Takahashi identities is as follows.
Assuming that the functional integral measure is invariant under gauge transformations, for a given set of fields $\Phi$, one has
\bea
\label{Eq:Ward1}
\delta Z[J]=\delta \int {\cal D}\hat{\Phi} e^{-(S[\hat{\Phi}]+\int J \hat{\Phi})}=0,
\eea
where $Z[J]$ is the partition function, $\delta$ denotes an infinitesimal gauge transformation, and note that we have introduced the fluctuating fields as $\hat{\Phi}$. Using the notation of the effective action, $\Gamma[{\Phi}]\equiv -\log Z[J]-\int J {\Phi}$, (\ref{Eq:Ward1})
leads to
\bea
\label{Eq:Ward2}
\langle \delta S[\hat{\Phi}] \rangle - \int_x \langle \delta \hat{\Phi}(x)\rangle \frac{\delta \Gamma}{\delta {\Phi}(x)}=0,
\eea
where $\langle ... \rangle$ refers to the average 
\bea
\langle ... \rangle=\int {\cal D}\hat{\Phi} (...) e^{-(S[\hat{\Phi}]+J\hat{\Phi})},
\eea
and $J=J[{\Phi}]$ is a function of $\Phi$ through the relation $J[{\Phi}]=-\delta \Gamma[{\Phi}]/\delta {\Phi}$. Note that $\langle \delta S \rangle\neq 0$ due to the explicit breaking of gauge symmetry via gauge fixing. In the FRG framework, a regulator term quadratic in the fields is added to the Lagrangian; thus, we have the substitution
\bea
\label{Eq:subs1}
S \quad \longrightarrow \quad S + \int_{xy} \Phi^\dagger(x) {\cal R}_k(x,y) \Phi(y).
\eea
Furthermore, the effective action generating 1PI diagrams changes as
\bea
\label{Eq:subs2}
\Gamma \quad \longrightarrow \quad \Gamma_k + \int_{xy} {\Phi}^\dagger(x) {\cal R}_k(x,y) {\Phi}(y),
\eea
where $\Gamma_k$ obeys the flow equation (\ref{Eq:floweq}). Applying the substitutions (\ref{Eq:subs1}) and (\ref{Eq:subs2}) on the Ward-Takahashi identity (\ref{Eq:Ward2}), one arrives at the modified Ward-Takahashi identity:
\bea
\label{Eq:mWTI}
\!\!\!\!\!\! \langle \delta S[\hat{\Phi}]\rangle&-&\int_x \langle \delta \hat{\Phi}(x)\rangle \frac{\delta \Gamma_k}{\delta {\Phi}(x)}=\nonumber\\
&-&\langle \delta\int_{xy} \hat{\Phi}^\dagger(x){\cal R}_k(x,y)\hat{\Phi}(y)\rangle+\nonumber\\
&&\!\!\!\!\!\!\!\!\!\int_x \langle \delta \hat{\Phi}(x)\rangle \frac{\delta}{\delta {\Phi}(x)}\int_{yz} {\Phi^\dagger}(y){\cal R}_k(y,z){\Phi}(z).
\eea
This identity can be considered as a master equation for individual (modified) Ward-Takahashi identities, which can be obtained by projecting both sides onto various operators. For example, if the regulator is missing and thus the right-hand side of (\ref{Eq:mWTI}) is zero, a projection onto $\sim A_i$ leads  to the absence of radiative corrections of the longitudinal photon, or $\sim \phi^{\dagger a}\phi^b$ shows the identity $Z_{e}=Z_{\phi}$. In Sec. IV, we discuss in detail how these identities are modified due to the regulator terms.

\section{Renormalization group analysis}

\subsection{Flow equations}

Now we are in a position to determine the $\gamma_{k,2}^{(i)}$ eigenvalues. First, we assume a spacetime dependent expectation value for $\sigma^a$, which covers both cases (\ref{Eq:bg1}) and (\ref{Eq:bg2}). As announced already, we set the freely adjustable field $\tilde{\sigma}$ (introduced via the gauge fixing term) as $\tilde{\sigma}^a=\sigma^a$. This choice considerably makes calculations easier, as the mixing term between $A_i$-$\pi^a$ reduces to $\sim \pi^a A_i \partial_i \sigma^a$. This also means that for a background where $\sigma^a \neq 0$, one does not need to go Fourier space in $\sigma^a$ to establish the quasilocal expansion of (\ref{Eq:flowlog}). One just evaluates all momentum space propagators in a constant background of $\sigma^a, \partial_i \sigma^a$ and obtain naturally $\Gamma_k$ as an integral of a local function in direct space.

Using the notation $\sigma^a\sigma^a=\sigma^2$, the second derivative matrix $\Gamma_{k,2}$ can be constructed from (\ref{Eq:gammaansatz}), and up to ${\cal O}\big((\partial_j \sigma)^2\big)$, the eigenvalues in Fourier space turn out to be the following:
\begin{widetext}
\bea
\label{Eq:eigen}
\gamma_{k,2}^{(1)}(q)&=&Z_{A,k}q^2/\xi_k+Z_ke^2\sigma^2-\frac{4e^2 Z_k^2(\hat{q}_j \partial_j \sigma)^2}{q^2(Z_k-Z_{A,k}/\xi_k)+Z_k(m_k^2+Z_k\lambda_k\sigma^2/6+\xi_kZ_k e^2\sigma^2/Z_{A,k}-e^2\sigma^2)},  \nonumber\\
\gamma_{k,2}^{(2)}(q)&=&Z_{A,k}q^2+Z_ke^2\sigma^2-\frac{4e^2Z_k^2\Big((\partial_j \sigma)^2-(\hat{q}_i\partial_j \sigma)^2\Big)}{q^2(Z_k-Z_{A,k})+Z_k(m_k^2+Z_k\lambda_k\sigma^2/6+\xi_kZ_k e^2\sigma^2/Z_{A,k}-e^2\sigma^2)}, 
\nonumber\\
\gamma_{k,2}^{(3)}(q)&=&Z_{A,k}q^2 + Z_ke^2\sigma^2, \hspace{1.75cm} [\mult: d-2] \nonumber\\
\gamma_{k,2}^{(d+1)}(q)&=&Z_k(q^2+m_k^2+Z_k\lambda_k\sigma^2/2), \nonumber\\
\gamma_{k,2}^{(d+2)}(q)&=&Z_k(q^2+m_k^2+Z_k\lambda_k\sigma^2/6), \hspace{0.45cm} [\mult: 2(N-1)] \nonumber\\
\gamma_{k,2}^{(d+2N)}(q)&=&Z_k(q^2+m_k^2+Z_k\lambda_k\sigma^2/6+\xi_k\frac{Z_k e^2}{Z_{A,k}}\sigma^2)+\frac{4e^2Z^2_k(\partial_j \sigma)^2}{q^2(Z_k-Z_{A,k})+Z_k(m_k^2+Z_k\lambda_k\sigma^2/6+\xi_kZ_k e^2\sigma^2/Z_{A,k}-e^2\sigma^2)} \nonumber\\
&+&[4e^2Z^2_k(q_j\partial_j \sigma)^2 (\xi_k^{-1}-1)Z_{A,k}]\times[q^2(Z_k-Z_{A,k})+Z_k(m_k^2+Z_k\lambda_k\sigma^2/6+\xi_kZ_k e^2\sigma^2/Z_{A,k}-e^2\sigma^2)]^{-1}\nonumber\\
&\times&[q^2(Z_k-Z_{A,k}/\xi_k)+Z_k(m_k^2+Z_k\lambda_k\sigma^2/6+\xi_kZ_k e^2\sigma^2/Z_{A,k}-e^2\sigma^2)]^{-1},\nonumber\\
\gamma_{k,2}^{(d+2N+1)}(q)&=&q^2+\xi_k\frac{Z_k}{Z_{A,k}}e^2\sigma^2, \hspace{1.75cm} [\mult: 2]
\eea
\end{widetext}
where $\hat{q}_i=q_i/|q|$. The first three eigenvalues originate from the timelike, longitudinal, and transverse polarizations of the photon, respectively. The fourth and fifth terms are the heavy and light (Nambu-Goldstone) modes, while the sixth is associated with the only direction of the $\pi^a$ field that receives an additional mass contribution via the gauge fixing term. The last term is the inverse ghost propagator. 
 
We still need to define the regulator matrix ${\cal R}_k$. The next subsection is devoted for discussions on the appropriate choice, but at the moment we just let it be a Litim-type function in every sector. Precisely speaking, the ${\cal R}_k$ regulator matrix is defined such that it diagonalizes as $\Gamma_{k,2}$, and (in Fourier space) a function $R_k^{(i)}(q)\equiv Z_k^{(i)}R_k(q)\equiv Z_k^{(i)}(k^2-q^2)\Theta(k^2-q^2)$ is associated to each eigenmode, where $Z_k^{(i)}$ is the coefficient of $q^2$ in the $i$th mode in ({\ref{Eq:eigen}). Note that, we will neglect the effect of $\tilde{\partial}_k$ on the $Z_k^{(i)}$ coefficients; it produces next-to-leading order contributions of the flows in terms of $\lambda_k$ and $e_k^2$. Having that in mind, one uses 
\bea
\partial_k R_k^{(i)} \approx 2kZ_k^{(i)}\Theta(k^2-q^2), 
\eea
then evaluates the integral in (\ref{Eq:flowlog}) (note that the ghosts appear with a negative sign due to their Grassmannian nature), and after expanding the obtained expression in $\sigma$ and $\partial_j \sigma$, compares the result with (\ref{Eq:gammaansatz}) to identify the operators given in (\ref{Eq:bg1}) and ({\ref{Eq:bg2}). One is then provided with the following flow equations:
\begin{subequations}
\bea
\label{Eq:flowm2}
k\partial_k (m_k^2Z_k)&=&-\frac{2}{3d}\frac{\Omega_d}{k^{2-d}}\Big[3(d-1)e_k^2+(N+1)\lambda_k\Big]Z_k, \nonumber\\ \\
\label{Eq:flowlZ2}
k\partial_k(\lambda_k Z_k^2)&=&\frac{4}{3d} \frac{\Omega_d}{k^{4-d}}\Big[18(d-1)e_k^4+6\xi_k e_k^2\lambda_k \nonumber\\
&+&(N+4)\lambda_k^2\Big]Z_k^2, \\
\label{Eq:flowZ}
k\partial_kZ_k&=&\frac{8}{d(d-2)} \frac{\Omega_d}{k^{4-d}}(d-1+\xi_k)e_k^2Z_k,
\eea
\end{subequations}
which have been evaluated for $m_k^2=0$ (i.e., in the critical point). We have also introduced the notation $\Omega_d=\int_\Omega (2 \pi)^{-d}\equiv 2/[(4\pi)^{d/2}\Gamma(d/2)]$ (here $\Gamma$ denotes Euler's gamma function). Combining (\ref{Eq:flowlZ2}) with (\ref{Eq:flowZ}) we get
\bea
\label{Eq:flowlk}
k\partial_k \lambda_k&=&\frac{4}{3d(d-2)}\frac{\Omega_d}{k^{4-d}}\Big[18(d-1)(d-2)e_k^4 \nonumber\\
&+&6e_k^2\lambda_k(2-2d-(4-d)\xi_k)\nonumber\\
&+&\lambda_k^2(d-2)(N+4)\Big].
\eea
The latter shows that the change in the self-coupling depends on the gauge fixing parameter unless the dimension is set to $d=4$. This peculiar statement originates from the fact that gauge symmetry cannot be maintained due the cutoff-type regularization that FRG enforces us to apply by construction.  This should not imply that the flow of $\lambda_k$ is arbitrary, as $\xi_k$ will be determined by a consistency relation between the flow equation and the applied ansatz for $\Gamma_k$.

In the following, we evaluate (\ref{Eq:flowlog}) in a classical field where the only nonvanishing component is $A_i(x)$. Since we are looking for the operator of (\ref{Eq:bg3}), it is not necessary to list all eigenvalues. Contributions arise from $N$ identical copies of $\sigma^a-\pi^a$ sectors that mix the following way:
\bea
\label{Eq:sigmapimix}
\Gamma^{\sigma^a\pi^a}_{k,2}(x,y)&=&Z_k
\left(\begin{matrix} -\partial^2+m_k^2+e^2A_i^2 & -e\partial_i A_i-2eA_i\partial_i \\ e\partial_i A_i+2eA_i\partial_i & -\partial^2+m_k^2+e^2A_i^2 \end{matrix}\right)\nonumber\\
&\times& \delta (x-y).
\eea
We may calculate the eigenvalues $\gamma^{(i)}_{k,2}$ as before, or just simply make use of the identity $\Tr \log = \log \det$. Plugging (\ref{Eq:sigmapimix}) into (\ref{Eq:floweq2}) we get
\bea
\label{Eq:gaugeflow}
\partial_k&&\!\!\!\!\!\!\!\Gamma_k|_A=\frac{N}{2} \tilde{\partial}_k \int \log [Z_k^2(-\partial^2+m_k^2+e^2A_i^2+R_k)^2 \nonumber\\
&&\hspace{2.5cm}+Z_k^2(e\partial_iA_i+2eA_i\partial_i)^2] \nonumber\\
&=&\frac{N}{2} \tilde{\partial}_k\bigg[\int \log[Z_k^2(-\partial^2+m_k^2+R_k)^2] \nonumber\\
&+&\int \log[1+2e^2(-\partial^2+m_k^2+R_k)^{-1}A_i^2] \nonumber\\
&+&\int \log[1+e^2[(-\partial^2+m_k^2+R_k)^{-1}(\partial_iA_i+A_i\partial_i)]^2\bigg], \nonumber\\
\eea
where we have omitted the functional unit matrices (delta functions) for the sake of compactness. Note that, the regulator is again constructed in the same fashion as above (i.e., associating individual regulators to eigenmodes); it turns out to be just adding $Z_k R_k$ in the diagonal components of (\ref{Eq:sigmapimix}). The first term in (\ref{Eq:gaugeflow}) does not contribute to the flow of $A_i$ dependent operators, thus can be discarded. The second and third terms can be expanded up to ${\cal O}(A^2)$ and evaluated readily in Fourier space:
\bea
\label{Eq:gaugeflowb}
\partial_k&& \Gamma_k|_A=\frac{Ne^2}{2}\int_p A_i(p)A_j(-p)\tilde{\partial}_k\Bigg[\int_q \frac{2\delta_{ij}}{q^2+m_k^2+R_k(q)}\nonumber\\
&&-\int_q \frac{(p+2q)_i(p+2q)_j}{[q^2+m_k^2+R_k(q)][(q+p)^2+m_k^2+R_k(q+p)]}\Bigg]\nonumber\\
&&+{\cal O}(A^4).
\eea
For $m_k^2=0$, the integrals combine into the following form in real space (see details in Appendix A):
\bea
\label{Eq:gaugeflow2}
\partial_k \Gamma_k|_A&=&\frac12 \Bigg(-\frac{4Ne^2(d-2)}{d(d+2)} \frac{\Omega_d}{k^{3-d}} \Bigg) \int_x A^2_i(x)\nonumber\\
&+&\frac12\Bigg(-\frac{8Ne^2}{d(d+2)}\frac{\Omega_d}{k^{5-d}}\Bigg) \nonumber\\
&\times&\int_x A_i(x) \Big(-\partial^2\delta_{ij}+\partial_i\partial_j \frac{d-2}{2}\Big) A_j(x)\nonumber\\
&+&{\cal O}(A^4,\partial^4 A^2).
\eea
First we notice that a mass term appeared for the photon, which was not present in (\ref{Eq:gammaansatz}). By adding the term $\frac12 \int_x m_{A,k}^2 A_i^2(x)$ to (\ref{Eq:gammaansatz}) we get
\bea
\label{Eq:mkAflow}
\partial_k m_{A,k}^2=-\frac{4Ne^2(d-2)}{d(d+2)}\frac{\Omega_d}{k^{3-d}}.
\eea
The appearance of the photon mass is again related to the breaking of gauge symmetry by the regulator. It is important to stress that, if at the UV scale it is adjusted to
\bea
\label{Eq:mkAbound}
m_{\Lambda,A}^2=-\frac{4Ne^2}{d(d+2)}\Omega_d\Lambda^{d-2},
\eea
then at $k=0$ it disappears as it should. Therefore, we can think of this phenomenon as an artifact of the formalism, which, after taken care of at the UV scale, does not have any physical effect in the infrared.

The second observation is that the flowing component of the gauge propagator is not transverse, as one would expect from perturbation theory (note that in perturbation theory no fluctuations affect the longitudinal part, as opposed to what we obtain here). The reason is the same as above; the regulator also violates the Ward-Takahashi identities, as we discussed in Sec. II (see also Sec. IV. in detail). One observes that consistency between the ansatz (\ref{Eq:gammaansatz}) and the flow equation (\ref{Eq:gaugeflow2}) requires the choice
\bea
\label{Eq:xicons}
\xi_k=2/(4-d),
\eea
which should also be used in (\ref{Eq:flowlk}). (\ref{Eq:xicons}) is inapplicable for $d=4$, in which case consistency requires $\xi_k$ to flow as the gauge wave function renormalization: $\xi_k=\const \times Z_{A,k}$ (choice of the constant is completely arbitrary). This is in accordance with perturbation theory.

Finally, we identify the flow of the photon wave function renormalization as
\bea
\label{Eq:ZAflow}
\partial_k Z_{A,k}=-\frac{8Ne^2}{d(d+2)}\frac{\Omega_d}{k^{5-d}}.
\eea
Since the connection between the bare and flowing charges is $e_k^2=e^2/Z_{A,k}$ (note that $e_\Lambda^2\equiv e^2$), we get
\bea
\label{Eq:flowe2k}
\partial_k e_k^2=\frac{8Ne_k^4}{d(d+2)}\frac{\Omega_d}{k^{5-d}}.
\eea

\subsection{Role of the regulator}

In the previous subsection we evaluated the flow equation by diagonalizing the matrix $\Gamma_{k,2}+{\cal R}_k$. First, we calculated the eigenvalues of $\Gamma_{k,2}$ and only after that defined the matrix ${\cal R}_k$ such that each eigenvalue received a contribution of a Litim-type function, i.e.,
\bea
R_k^{(i)}(q)=Z_k^{(i)}R_k(q)=Z_k^{(i)}(k^2-q^2)\Theta(k^2-q^2). 
\eea
It is easy to see that even if we decide to fix the shape regulator by the Litim function $R_k(q)$, a separate choice of regularization is also legitimate. By this we mean that we could have added the following term to the classical action in advance to diagonalizing $\Gamma_{k,2}$:
\bea
\label{Eq:reg2}
\int&&\!\!\!\!\!\!\!\Phi^\dagger {\cal R}_k \Phi \equiv\nonumber\\
&&\frac{Z_{A,k}}{2} \int_q A_i(q)R_k(q) \left(\delta_{ij}+\frac{q_i q_j} {q^2}(1-\xi^{-1}_k)\right) A_j (-q)\nonumber\\
&+&Z_{\phi,k}\int_q \phi^{\dagger a}(q)R_k(q)\phi^a(-q)+\int_q c^*(q)R_k(q)c(-q),\nonumber\\
\eea
which would have led the flow equation to
\bea
\label{Eq:flowlog2}
\partial_k \Gamma_k^{\Rtwo} = \frac12 \tilde{\partial}_k \int_x \int_{|q|<k} \sum_{i=1}^{2N+d+2} \log [\gamma_{k,2}^{(i)}(k)],
\eea
instead of 
\bea
\label{Eq:flowlogb}
\partial_k \Gamma_k^{\Rone} = \frac12 \tilde{\partial}_k\int_x \int_{|q|<k} \!\!\!\!\!\sum_{i=1}^{2N+d+2} \log [\gamma_{k,2}^{(i)}(q)-Z_k^{(i)}(q^2-k^2)], \nonumber\\
\eea
the latter being a direct consequence of (\ref{Eq:flowlog}) (the superscripts of $\Gamma_k$ distinguish different regulators). The difference between the two approaches is that while in (\ref{Eq:flowlogb}) the regulator only changes the Gaussian quadratic momentum dependence to $k^2$, in (\ref{Eq:flowlog2}) it completely eliminates $q$, and everywhere replaces it with $k$.

Looking at ({\ref{Eq:eigen}) one observes that if $\sigma \neq 0, \partial_j \sigma=0$, then there is no additional $q$ dependence apart from the Gaussian part, and all eigenvalues (\ref{Eq:eigen}) take the form of $\gamma_{k,2}^{(i)}=Z_k^{(i)}q^2+\const$, therefore
\bea
\label{Eq:qdep}
\gamma_{k,2}^{(i)}(q)-Z_k^{(i)}(q^2-k^2)=\gamma_{k,2}^{(i)}(k).
\eea
This case is of no interest as the two regularization procedures turn out to be equivalent. However, when $\partial_j \sigma \neq 0$, the two sides of (\ref{Eq:qdep}) differ, and so does the corresponding prediction of the flow equation, that is the $k$ dependence of the scalar wave function renormalization $Z_k$. At this point one has to decide which choice is more reliable.

Since only the flow of $Z_k$ differs in the two cases, formally we can consider $\partial_j \sigma \neq 0$ but set $\sigma=0$ to obtain $\partial_k Z_k$. The eigenvalues of (\ref{Eq:eigen}) that contribute are
\begin{subequations}
\label{Eq:eigen2}
\bea
\gamma_{k,2}^{(1)}(q)&=&Z_{A,k}q^2/\xi_k-\frac{4e^2 Z_k^2(\partial_j \sigma)^2}{(Z_k-Z_{A,k}/\xi_k)q^2}\cos^2 \theta,  \nonumber\\
\label{Eq:eigen2a} \\
\gamma_{k,2}^{(2)}(q)&=&Z_{A,k}q^2-\frac{4e^2Z_k^2(\partial_j \sigma)^2}{(Z_k-Z_{A,k})q^2}(1-\cos^2\theta), \nonumber\\
\label{Eq:eigen2b} \\
\gamma_{k,2}^{(d+2N)}(q)&=&Z_kq^2+\frac{4e^2Z^2_k(\partial_j \sigma)^2}{(Z_k-Z_{A,k})q^2}\nonumber\\
&-&\frac{4e^2 Z^2_k(1-\xi^{-1}_k)Z_{A,k}(\partial_j \sigma)^2}{(Z_k-Z_{A,k})(Z_k-Z_{A,k}/\xi_k)q^2}\cos^2\theta. \nonumber\\
\label{Eq:eigen2c}
\eea
\end{subequations}
Here we denoted the angle between $q_j$ and $\partial_j \sigma$ by $\theta$. Plugging (\ref{Eq:eigen2a}),  (\ref{Eq:eigen2b}), and (\ref{Eq:eigen2c}) into (\ref{Eq:flowlog2}) and (\ref{Eq:flowlogb}), we arrive at a series representation of $\partial_k \Gamma_k$ in terms of $\partial_i \sigma$. Detailed calculations can be found in Appendix B; here we just state that by taking the leading order term in the obtained expansions we get
\begin{subequations}
\label{Eq:waveR}
\bea
\label{Eq:waveR1}
\partial_k \log Z_k^{\Rone}&=&\frac{8e_k^2}{d(d-2)} \frac{\Omega_d}{k^{5-d}}(d-1+\xi_k), \nonumber\\ \\
\label{Eq:waveR2}
\partial_k \log Z_k^{\Rtwo}&=&\frac{16e_k^2}{d^2} \frac{\Omega_d}{k^{5-d}}(d-1+\xi_k). \nonumber\\
\eea
\end{subequations}
By obtaining (\ref{Eq:waveR1}) we have reproduced (\ref{Eq:flowZ}), but we see that the predictions for $Z_k$ differ in the two regularization procedures. Comparing (\ref{Eq:waveR1}) and (\ref{Eq:waveR2}) we get $\partial_k \log Z_k^{\Rone}/\partial_k\log Z_k^{\Rtwo}=d/2(d-2)$. It is worth noting that for $d=4$, $Z_k^{\Rone}=Z_k^{\Rtwo}$, which is no surprise as in $d=4$ the $\beta$ functions are universal. For $d<4$ ($d>4$) we get $Z_k^{\Rone}>Z_k^{\Rtwo}$ ($Z_k^{\Rone}<Z_k^{\Rtwo}$).

The fact that two regularizations lead to different predictions is clearly due to the truncation of the effective action (\ref{Eq:gammaansatz}). If one was able to solve the flow equation exactly, there would be no regulator dependence whatsoever, but (\ref{Eq:waveR1}) and (\ref{Eq:waveR2}) show that in the current approximation one has to make a choice regarding the two regularization schemes. We decide by looking at the convergence properties of the series representations in question [see (\ref{Eq:R1b}) and (\ref{Eq:R2b}) in Appendix B]. One observes that while both cases lead to alternating series, in case of R1, the absolute value of the coefficients is decreasing as we proceed to higher orders, as opposed to that of R2. This shows that if one is to take a finite truncation [which is certainly what we are doing in (\ref{Eq:gammaansatz}) by keeping only the leading order term], the former (i.e., regulator R1) should be taken as a better approximation since one neglects terms that gradually carry less contribution. 

\subsection{Fixed points}

It is a four-dimensional space of coupling constants and mass parameters ($\lambda_k, e_k^2, m_k^2, m_{A,k}^2$) in which one is looking for fixed points. We emphasize that it is not necessary to determine the complete map of trajectories of the system in order to find fixed points, as for this purpose one may use approximate flow equations that are valid when $k\rightarrow 0$. First of all, the photon mass $m_{A,k}^2$ is clearly nonphysical, but we saw in the previous subsection that it dies out as $k\rightarrow 0$. Therefore, for a fixed point analysis it is safe to set $m_{A,k}^2=0$ and analyze the flows accordingly (as we assumed it implicitly when deriving the RG equations of the couplings and wave function renormalizations). Because of a similar argument, one may choose $m_k^2=0$, as it corresponds to the temperature parameter. This does not mean that the flow of $m_k^2$ is neglected, to the contrary, it does flow, but the initial conditions are considered to be set in a way that it approaches zero when $k \rightarrow 0$ (this signals the critical point). By these considerations, one is looking for the fixed point structure in the $(\lambda_k, e_k^2$) plane via the following equations [see (\ref{Eq:flowlk}) and (\ref{Eq:flowe2k})]:
\begin{subequations}
\bea
\label{Eq:flowlambdanoxi}
k\partial_k \lambda_k&=&\frac{4}{3d(d-2)} \frac{\Omega_d}{k^{4-d}}[18(d-1)(d-2)e_k^4 \nonumber\\
&-&12de_k^2\lambda_k+(d-2)(N+4)\lambda_k^2], \\
\label{Eq:flowe2k2}
\partial_k e_k^2&=&\frac{8Ne_k^4}{d(d+2)} \frac{\Omega_d}{k^{4-d}},
\eea
\end{subequations}
where we used the consistency condition $\xi_k(4-d)=2$ in $(\ref{Eq:flowlambdanoxi})$. The $d=4$ case is special as there no $\xi_k$ dependence appears, and we directly get
\begin{subequations}
\bea
\label{Eq:l4}
k\partial_k \lambda_k|_{d=4}&=&\frac{54e_k^4-18e_k^2\lambda_k+(N+4)\lambda_k^2}{24\pi^2}, \\
\label{Eq:e24}
\partial_k e_k^2|_{d=4}&=&\frac{Ne_k^4}{24\pi^2}.
\eea
\end{subequations}
\begin{figure}[t]
\includegraphics[bb = 0 350 515 255,scale=0.35,angle=270]{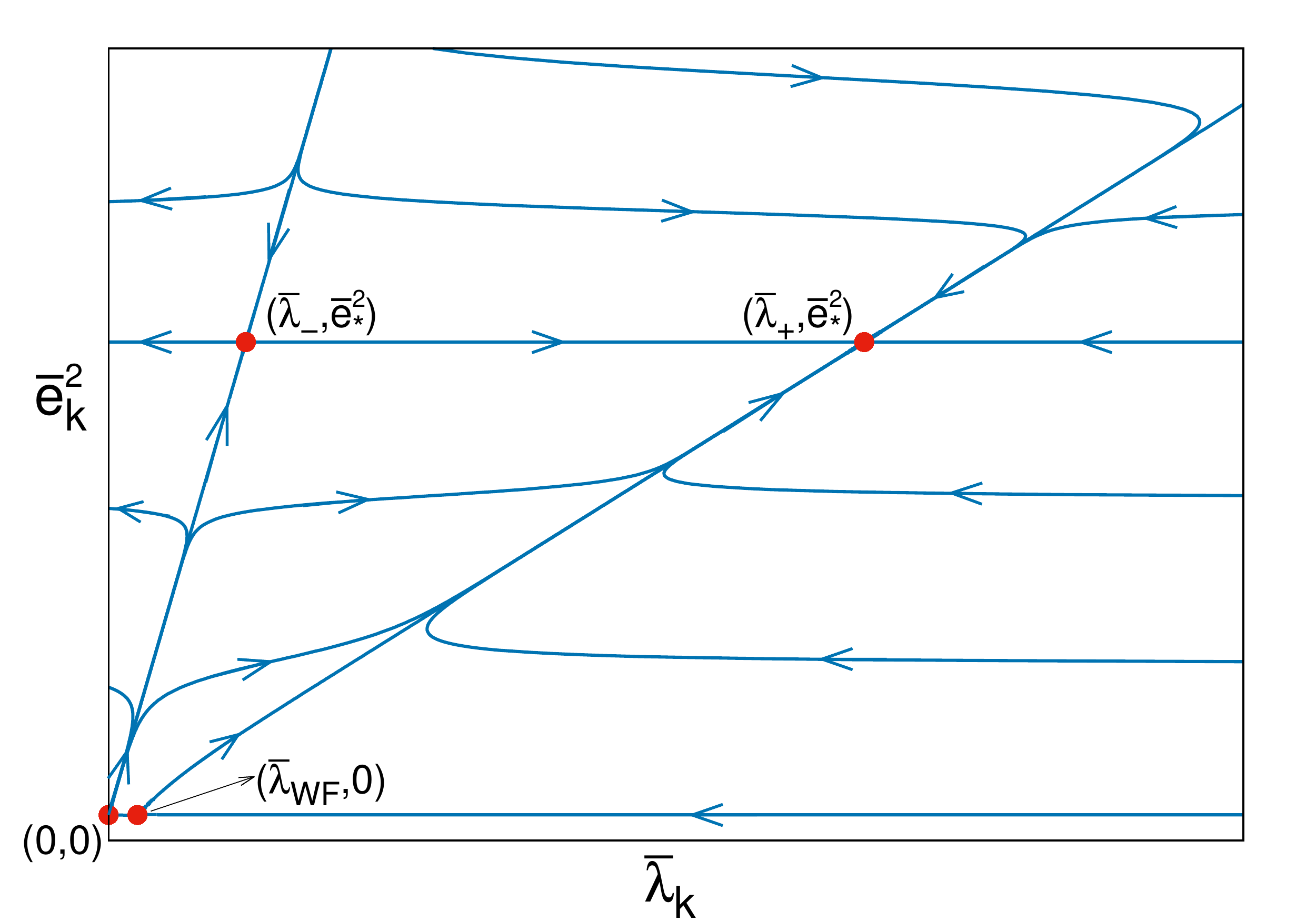}
\caption{Structure of the renormalization group flows in $d=3$. Two charged fixed points appear for every $N$. Arrows point from UV to IR.}
\end{figure}  
Note that (\ref{Eq:l4}) and (\ref{Eq:e24}) are the standard results of perturbation theory. Since fixed points are found in terms of dimensionless variables $\bar{\lambda}_k=\lambda_k k^{d-4}$, $\bar{e}^2_k=e_k^2 k^{d-4}$, Eqs. (\ref{Eq:flowlambdanoxi}) and (\ref{Eq:flowe2k2}) should be rewritten as
\bea
\label{Eq:betal}
\beta_\lambda&\equiv& k\partial_k \bar{\lambda}_k=(d-4)\bar{\lambda}_k+\frac{4\Omega_d}{3d(d-2)} \nonumber\\
&&\hspace{-1cm}\times[18(d-1)(d-2)\bar{e}_k^4-12d\bar{e}_k^2\bar{\lambda}_k+(d-2)(N+4)\bar{\lambda}_k^2], \nonumber\\ \\
\label{Eq:betae2}
\beta_{e^2}&\equiv& k\partial_k \bar{e}_k^2=(d-4)\bar{e}_k^2+\frac{8N\Omega_d}{d(d+2)}\bar{e}_k^4.\nonumber\\
\eea
For uncharged fixed points one solves (\ref{Eq:betal}) with $e_k^2\equiv 0$ and obtains
\bea
\bar{\lambda}_{\textrm{G}}=0, \qquad \bar{\lambda}_{\textrm{WF}}=\frac{3d(4-d)}{4(N+4)\Omega_d}.
\eea
Note that, $\bar{\lambda}_{\textrm{G}}$ and $\bar{\lambda}_{\textrm{WF}}$ correspond to the Gaussian and Wilson-Fisher fixed points, respectively. For the latter, one has to have $d<4$ due to stability reasons, which remains as a necessary condition for the existence of charged fixed points, as $e_k^2>0$. (\ref{Eq:betae2}) shows that charged fixed points may exist on the line of 
\bea
\label{Eq:e2fixed}
\bar{e}^2_k= \bar{e}^2_*\equiv d(d+2)(4-d)/8N\Omega_d,
\eea
which indeed appear if one solves (\ref{Eq:betal}) with (\ref{Eq:e2fixed}) as a constraint:
\bea
\bar{\lambda}_\pm=3d\frac{(4-d)(2d(d+2)+(d-2)N)\pm \sqrt{\Delta}}{8(d-2)N(N+4)\Omega_d},
\eea
where
\bea
\Delta&=&(d^2-6d+8)^2N^2\nonumber\\
&-&2(d-4)^2(d-2)(d+2)\big(4+(d-3)d(d+2)\big)N\nonumber\\
&-&4(d-4)^2(d+2)^2(d^2(2d-11)+16d-8).
\eea
In particular, in $d=3$ one has
\bea
\bar{\lambda}_{\textrm{G}}|_{d=3}=0, \qquad \bar{\lambda}_{\textrm{WF}}|_{d=3}=\frac{9}{4(N+4)\Omega_3}, \nonumber\\
\bar{\lambda}_{\pm}|_{d=3}=\frac{9(30+N\pm\sqrt{(20-N)^2+100})}{8 N(N+4)\Omega_3},
\eea
with
\bea
\bar{e}^2_*|_{d=3}=\frac{15}{8N\Omega_3}.
\eea
At $N=1$ we reproduce the results of \cite{fejos16}, where it has already been reported that this structure can describe the superconducting phase transition, see the UV$\rightarrow$IR flows in Fig. 1.  We also see in Fig. 2 that, as opposed to the standard results of the $\epsilon$ expansion, the current fixed point structure remains valid for every $N$.

\begin{figure}[t]
\includegraphics[bb = 0 370 495 255,scale=0.38,angle=270]{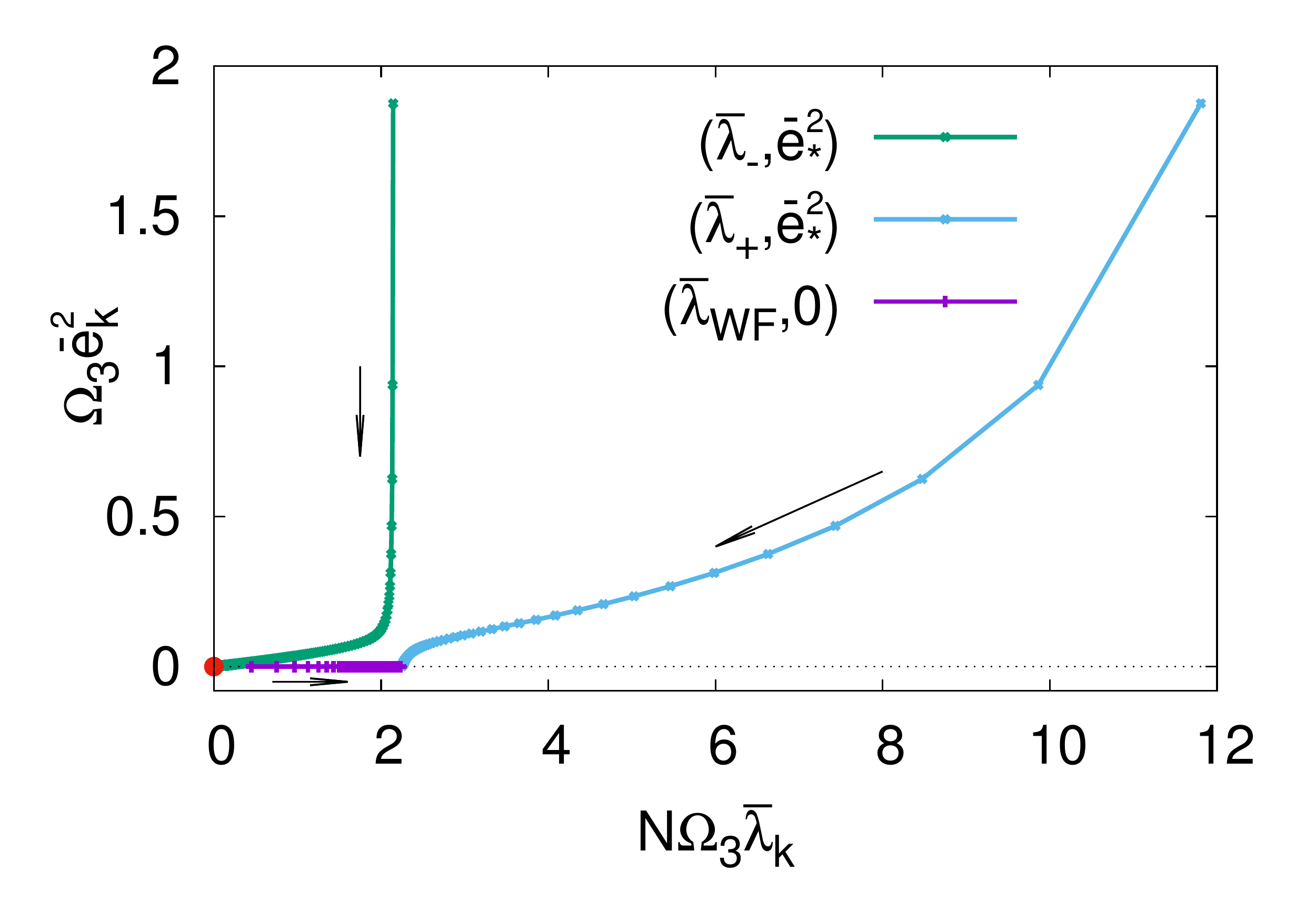}
\caption{Fixed points as a function of $N$ in $d=3$. Arrows point into directions where $N$ is increasing.}
\end{figure}  

The region of the parameter space that is attracted by the fixed point $(\lambda_+,\bar{e}^2_*)$ can be uniquely characterized by the Ginzburg-Landau parameter $\kappa^2_k\equiv \bar{\lambda}_k/6\bar{e}_k^2$. The RG flow of $\kappa^2_k$ is
\bea
k\partial_k \kappa^2_k = \frac{8\Omega_3}{15}\bar{e}_k^2\big(5(N+4)\kappa_k^2-(N+30)\kappa_k+5\big),
\eea
which shows that $\kappa_k$ has two fixed points (note that, $\kappa_k>0$, since $\lambda_k>0$ and $e^2_k>0$):
\bea
\kappa_{\pm}=\sqrt{\frac{30+N\pm \sqrt{(20-N)^2+100}}{10(N+4)}},
\eea
where $\kappa_+$ is stable, while $\kappa_-$ is unstable, see also Fig. 3. That is if $\kappa_k>\kappa_-$, the RG flows drive the system toward the IR fixed point $(\lambda_+,\bar{e}^2_*)$ and it undergoes a second order transition. However, if $\kappa_k<\kappa_-$, the transition cannot be continuous and presumably it is of first order. For $N=1$ we get $\kappa_-\approx 0.62/\sqrt2$, in accordance with \cite{fejos16} and in decent agreement with Monte Carlo simulations \cite{mo02}.

We close this section by drawing attention on the importance of the use of the R1 regulator. Had we chosen R2, the flow of the wave function renormalization $Z_k$ would have received a multiplicative factor of $2/3$ [see (\ref{Eq:waveR})]. This would have affected the flow of the self-coupling $\lambda_k$ and changed the positions of the fixed points. A short calculation leads to the conclusion that this change is so dramatic that the charged fixed points disappear if $N\leq N_{\crit}=68$. Since one expects the existence of these fixed points even for $N=1$, this shows the importance of optimizing the flows through the regulator choice, and in addition to the arguments presented in the previous subsection, backs the choice of regulator R1 instead of R2.

\begin{figure}[t]
\includegraphics[bb = 0 370 515 255,scale=0.355,angle=270]{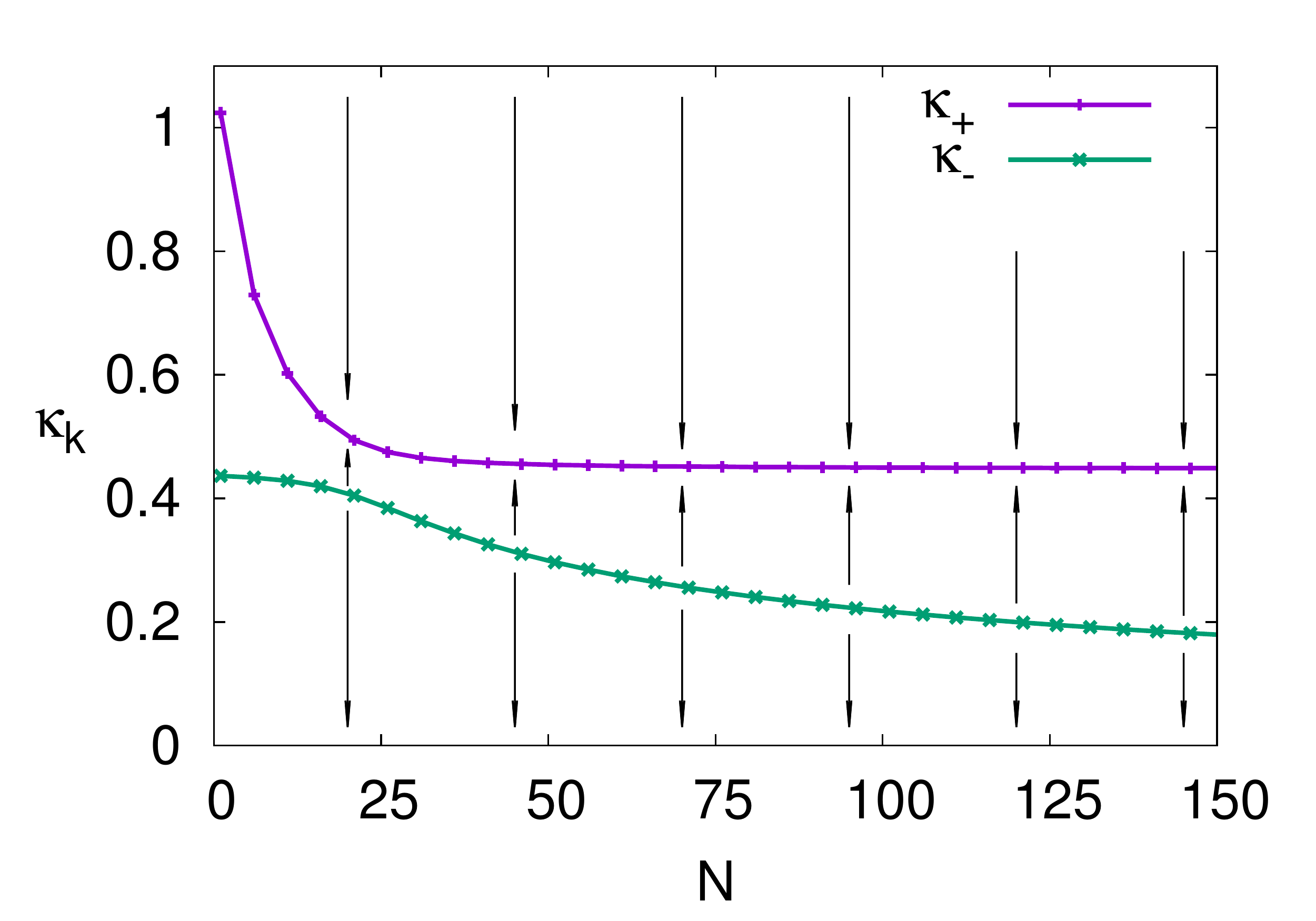}
\caption{Flow diagram of the Ginzburg-Landau parameter $\kappa_k$ in $d=3$. Arrows point from UV to IR. Above the line of $\kappa_{-}$ the phase transition is of second order.}
\end{figure}

\section{Evaluation of the modified Ward-Takahashi identities}

In this section we evaluate the modified Ward-Takahashi identities. One is interested in the anomalous contributions of the IR regulator, and the relationship between these identities and the flow equation. {\it A priori} there is no reason to expect that after truncating the effective action their compatibility (given in the full theory) remains \cite{gies12,igarashi16}. Also, we have made use of the equality $Z_{e,k}=Z_{\phi,k}$, which was put in the flow equations by hand, but might be violated due to the regulator. The master equation of the mWTI, Eq. (\ref{Eq:mWTI}), is therefore evaluated for two projections, $\sim A_i$ (gauge identity) and $\sim \phi^{\dagger a}\phi^b$ (scalar identity), respectively.

In what follows we restore the difference between $Z_{\phi,k}$ and $Z_{e,k}$, and use the basis of $(\phi^{\dagger a},\phi^a)$ instead of $(\sigma^a,\pi^a)$. The ansatz of $\Gamma_k$ [based on (\ref{Eq:gammaansatz0})]
is
\bea
\label{Eq:gammaansatz2}
\Gamma_k\!\!\!\!\!&&= \int_x \bigg[\frac{Z_{A,k}}{2}{A}_i\Big(-\partial^2 \delta_{ij}+\partial_i\partial_j (1-\xi_k^{-1})\Big){A}_j \nonumber\\
&+&Z_{\phi,k}\phi^{\dagger a} (-\partial^2+m_k^2)\phi^a+\frac{\lambda_kZ_{\phi,k}^2}{6}(\phi^{\dagger a} \phi^a)^2\nonumber\\
&-&iZ_{e,k}eA_i(\partial_i \phi^{\dagger a} \phi^a - \phi^{\dagger a} \partial_i \phi^a)+\frac{Z^2_{e,k}}{Z_{\phi,k}}e^2A_iA_i \phi^{\dagger a} \phi^a \nonumber\\
&+&\sqrt2e\partial_i A_i Z_{e,k}\tilde{\sigma}^a\Im\phi^a+\xi_k\frac{Z_{e,k}^2}{Z_{A,k}}e\tilde{\sigma}^a\tilde{\sigma}^b\Im\phi^a\Im\phi^b \nonumber\\
&+&c^*(-\partial^2+\xi_k\frac{\sqrt2 Z_{e,k}^2}{Z_{A,k}Z_{\phi,k}}e^2\tilde{\sigma}^a\Re\phi^a)c\bigg].
\eea

\subsection{The gauge identity}

First, we are looking for the renormalization of the gauge propagator; therefore, we need to project (\ref{Eq:mWTI}) onto the operator $\sim$$A_i$. One is free to work in the symmetric phase; thus,
we set $\tilde{\sigma}^a=0$, which completely decouples the ghost fields $(c^*,c)$. Let us begin with calculating the left-hand side of $(\ref{Eq:mWTI})$. It is convenient to work in Fourier space. First, one gets
\bea
\langle \delta S[\hat{\Phi}]\rangle &=&\frac{1}{2\xi} \langle \delta\int_x \big(\partial_i \hat{A}_i(x)\big)^2\rangle \nonumber\\
&=& \frac{i}{\xi} \int_p \theta(-p) p^2 p_i {A}_i(p),
\eea
and an almost identical term comes from the second term of the lhs of (\ref{Eq:mWTI}). One arrives at
\bea
\label{Eq:lhs}
\lhs = i\int_p \theta(-p)p^2p_i{A}_i(p)\left[\frac{1}{\xi}-\frac{Z_{A,k}}{\xi_k}\right].
\eea
Note that if the rhs of (\ref{Eq:mWTI}) was zero, we would get $\xi_k=\const \times Z_{A,k}$, which agrees with the expectations based on perturbation theory.

However, the rhs due to the regulator is nonzero, and the ${\cal{O}}({A}_i)$ piece of it is carried solely by scalar fluctuations. The corresponding term that has to be evaluated is the scalar regulator $\int Z_{\phi,k} \hat{\phi}^{\dagger a} R_k \hat{\phi}^a$, and using the transformation properties (\ref{Eq:trf}), we get
\bea
\langle \delta \int_{xy} &&Z_{\phi,k} \hat{\phi}^{\dagger a}(x) R_k(x,y) \hat{\phi}^a(y)\rangle =\nonumber\\
&&iZ_{\phi,k}e\int_p \theta(-p)\int_q \langle \hat{\phi}^{\dagger a}(q)\hat{\phi}^a(q+p)\rangle \times\nonumber\\
&&\hspace{2.8cm}\left[R_k(p+q)-R_k(q)\right].
\eea
Using this identity in the rhs of (\ref{Eq:mWTI}) we get
\bea
\label{Eq:rhs}
\rhs = -iZ_{\phi,k}e\int_p && \theta(-p)\int_q \Big(\langle \hat{\phi}^{\dagger a}(q-p)\hat{\phi}^a(q)\rangle_c\nonumber\\
&&-\langle \hat{\phi}^{\dagger a}(q)\hat{\phi}^a(q+p)\rangle_c\Big) R_k(q),
\eea
where the connected two-point function is
\bea
\langle \hat{\phi}^{\dagger a}(p')\hat{\phi}^b(p)\rangle_c&=&\langle \hat{\phi}^{\dagger a}(p')\hat{\phi}^b(p)\rangle \nonumber\\
&-&\langle \hat{\phi}^{\dagger a}(p')\rangle \langle\hat{\phi}^b(p)\rangle,
\eea
which has to be evaluated in the presence of a classical field $A_i$. From (\ref{Eq:gammaansatz2}), after inverting the corresponding two-point vertex, in the desired accuracy
\bea
\label{Eq:conn}
\langle \hat{\phi}^{\dagger a}(p')\hat{\phi}^b(p)\rangle_c|_A&=&\frac{1}{Z_{\phi,k}\big(p^2+R_k(p)\big)}\delta(p-p')\delta^{ab}\nonumber\\
&+& \frac{Z_{e,k}}{Z^2_{\phi,k}}\frac{eA_i(p-p')(p+p')_i}{\big(p^2+R_k(p)\big)\big(p'^2+R_k(p')\big)}\delta^{ab}\nonumber\\
&+&{\cal{O}}(A^2).
\eea
The easiest way to obtain (\ref{Eq:conn}) is to recall that for operators $\Gamma_{k,2}^{(0)}$ and $P$,
\bea
\label{Eq:inverse}
\!\!(\Gamma_{k,2}^{(0)}-P)^{-1}=(\Gamma_{k,2}^{(0)})^{-1}\Big(1+\sum_{n=1}^{\infty} \big(P(\Gamma_{k,2}^{(0)})^{-1}\big)^n\Big),
\eea
(also known as the Neumann series) and consider the propagator mixing as perturbation $P$. Note that, in (\ref{Eq:inverse}) multiplications have to be considered both in the functional and matrix sense.  Inserting (\ref{Eq:conn}) into (\ref{Eq:rhs}), at $m_k^2=0$, we get
\bea
\label{Eq:rhsfin}
\rhs&=&iNe^2\frac{Z_{e,k}}{Z_{\phi,k}}\int_p \theta(-p) A_i(p) \nonumber\\
&\times&\int_q \frac{2(2q+p)_iR_k(q)}{[q^2+R_k(q)][(q+p)^2+R_k(q+p)]}.
\eea 
A straightforward calculation (see Appendix A for details) leads to
\bea
\label{Eq:rhsfinal}
\rhs&=&i\Omega_d\frac{4Ne^2}{d(d+2)}\frac{Z_{e,k}}{Z_{\phi,k}}\int_p \theta(-p){A}_i(p) \nonumber\\
&\times&\left(k^{d-2}p_i-k^{d-4}p^2p_i +{\cal O}(p^5)\right) + {\cal O}(A^2).
\eea
First, one notices (similarly as in Sec. III) that there is no term in (\ref{Eq:lhs}) that would correspond to the first term in the bracket in (\ref{Eq:rhsfinal}). It shows the sign of a photon mass, which was left out from (\ref{Eq:gammaansatz2}) and thus has no correspondence in (\ref{Eq:lhs}). If one adds $\frac12 \int_x m_{A,k}^2 A_i^2(x)$ to (\ref{Eq:gammaansatz2}), then the mWTI tells that
\bea
m_{A,k}^2=-\frac{4Ne^2}{d(d+2)}\frac{Z_{e,k}}{Z_{\phi,k}}\Omega_d k^{d-2},
\eea
which, by assuming $Z_{e,k}=Z_{\phi,k}$, gives the same result as the flow equation (\ref{Eq:mkAflow}), including the boundary condition (\ref{Eq:mkAbound}). Second, matching (\ref{Eq:rhsfinal}) with (\ref{Eq:lhs}), the other term in the bracket of (\ref{Eq:rhsfinal}) shows that the following consistency condition has to be satisfied:
\bea
\label{Eq:mWTIfinal}
\frac{Z_{A,k}}{\xi_k}-\frac{1}{\xi}=\frac{4Ne^2}{d(d+2)}\frac{Z_{e,k}}{Z_{\phi,k}} \frac{\Omega_d}{k^{4-d}}.
\eea
Applying the operator $k\partial_k$ on (\ref{Eq:mWTIfinal}),
\bea
\label{Eq:mWTIfinalder}
\!\!\!\!\!\!\frac{k\partial_k Z_{A,k}}{\xi_k}-Z_{A,k}\frac{k\partial_k \xi_k}{\xi_k^2}&=&\frac{4Ne^2(d-4)}{d(d+2)}\frac{Z_{e,k}}{Z_{\phi,k}}\frac{\Omega_d}{k^{4-d}} \nonumber\\
&+&{\cal O}(e^4)
\eea
and we see that it is compatible with the corresponding flow equation (\ref{Eq:gaugeflow2}), provided $\xi_k$ satisfies the following. The choice $d=4$ shows that $\xi_k=\const \times Z_{A,k}$ (just as if the regulator was absent), while $d\neq 4$ leads to (\ref{Eq:ZAflow}) if
\bea
\xi_k = 2/(4-d)
\eea
is satisfied as a fixed point, and given that $Z_{e,k}=Z_{\phi,k}$. This is in an exact agreement with what we found when analyzing the compatibility of the flow equation and the ansatz of the effective action.

\subsection{The scalar identity}

Now we investigate the validity of the $Z_{e,k}=Z_{\phi,k}$ relation, which originally can be proven by projecting the WTI [lhs of (\ref{Eq:mWTI})] onto $\sim \phi^{\dagger a}\phi^b$. For this projection $\delta S$ gives no contribution, and the lhs of (\ref{Eq:mWTI}) is just the well-known Ward-Takahashi identity:
\bea
\label{Eq:lhs_b}
\lhs &=& i\int_p \theta(-p)\int_q \phi^{\dagger a}(q)\phi^b(q+p)\nonumber\\
&\times& \Big[e\Gamma^{ab}_k(q+p)-e\Gamma^{ab}_k(q)-p_i\Gamma_{i,k}^{ab}(p,q)\Big],
\eea
where $\Gamma^{ab}_k$ is the scalar inverse connected two-point function, and $\Gamma_{i,k}^{ab}$ is the $A^i$-$\phi^{\dagger a}$-$\phi^b$ vertex (in Fourier space). One reads off from the ansatz (\ref{Eq:gammaansatz2}) that (at $m_k^2=0$) $\Gamma_k^{ab}(q)=Z_{\phi,k}q^2\delta^{ab}$, $\Gamma_{i,k}^{ab}(p,q)=Z_{e,k}e(p+2q)_i\delta^{ab}$, which leads (\ref{Eq:lhs_b}) to
\bea
\label{Eq:leftfinal}
\lhs &=& -ie\int_p \theta(-p)\int_q \phi^{\dagger s}(q)\phi^s(q+p) \nonumber\\
&\times& (p^2+2p\cdot q) (Z_{e,k}-Z_{\phi,k}).
\eea
If there was no regulator, we would get $Z_{\phi,k}=Z_{e,k}$, but contributions may again appear from the rhs of (\ref{Eq:mWTI}). These anomalous terms can arise solely from the scalar regulator, and one has to again evaluate (\ref{Eq:rhs}), but now calculated in the presence of a classical field $\phi^a$ (and by setting the gauge field $A_i=0$). To simplify calculations, we work in R2 regularization, and require the freely adjustable field $\tilde{\sigma}^a$ in the gauge fixing condition to be zero, $\tilde{\sigma}^a=0$, even though throughout the paper we set it equal to $\sigma^a$.

The reason is that by adding (\ref{Eq:gf}) into the Lagrangian, one implicitly assumes that $\phi^a$ is real, and introduces asymmetry between $\sigma^a \leftrightarrow \pi^a$, i.e., breaks orthogonal invariance. If $\pi^a$ is set to zero, this makes no problem, but a direct calculation of the flow of $Z_{e,k}$ (this is to be done in Appendix C) requires nonzero expectation value for both $\sigma^a$ and $\pi^a$. We thus work with $\tilde{\sigma}^a=0$, which does not introduce asymmetry and maintains orthogonal invariance, and this way we can indeed confirm compatibility of the flow equation and the mWTI. We note that these restrictions should not affect the conclusions to be drawn at the end of this subsection.

For indices that correspond to different directions than that of the classical field $\phi^s$, we have (at $m_k^2=0$)
\bea
\label{Eq:conn2phi}
\langle\hat{\phi}^{\dagger a}(p')\hat{\phi}^b(p)\rangle_c|_\phi&=&\frac{1}{Z_{\phi,k}\big(p^2+R_k(p)\big)}\delta(p-p')\delta^{ab}\nonumber\\
&+&\frac{\lambda_k}{3} \frac{\int_l \phi^{\dagger s}(l-p)\phi^s(l-p')}{\big(p'^2+R_k(p')\big)\big(p^2+R_k(p)\big)}\delta^{ab}\nonumber\\
&+&{\cal O}(\phi^4).
\eea
These propagators do not give any contribution when evaluating the traces in (\ref{Eq:rhs}), which can be seen by performing a simultaneous, $q \rightarrow -q$, $l \rightarrow l+p$ variable change in the second term of the bracket of (\ref{Eq:rhs}). For the direction $s$ that corresponds to the symmetry breaking, a more careful treatment is necessary. First, one notes that the coefficient of $\sim \phi^{\dagger s}\phi^s$ receives an extra factor of $2$, but more importantly, the corresponding $\hat{\phi}^s$ fluctuating field mixes with the gauge fields. For calculating the (connected) two-point function in question, a $(d+2)\times (d+2)$ submatrix of $\delta^2\Gamma_k/\delta \Phi^\dagger \Phi$ needs to be inverted (spanned by the fluctuating fields $\hat{A}_i$, $\hat{\phi}^s$ and its conjugate $\hat{\phi}^{\dagger s}$). Treating the propagator mixing as perturbation, using (\ref{Eq:inverse}), we get
\bea
\label{Eq:conn2phi}
\langle\hat{\phi}^{s\dagger}(p')&&\!\!\!\!\!\!\!\hat{\phi}^s(p)\rangle_c|_\phi=\frac{1}{Z_{\phi,k}\big(p^2+R_k(p)\big)}\delta(p-p') \nonumber\\
&+&\frac{2\lambda_k}{3} \frac{\int_l \phi^{\dagger s}(l-p)\phi^s(l-p')}{\big(p'^2+R_k(p')\big)\big(p^2+R_k(p)\big)} \nonumber\\
&+&\frac{4e^2}{Z_{A,k}} \frac{Z_{e,k}^2}{Z_{\phi,k}^2}\int_l f(p,p';l)\phi^{\dagger s}(l-p)\phi^s(l-p')\nonumber\\
&+&{\cal O}(\phi^4).
\eea
where
\bea
f(p,p';l)&=&\frac{l^2(p\cdot p')-(l\cdot p)(l\cdot p')}{l^2(l^2+R_k(l))(p^2+R_k(p))(p'^2+R_k(p'))} \nonumber\\
&+&\frac{\xi_k}{4}\frac{(l^2-2l\cdot p)(l^2-2l\cdot p')}{l^2(l^2+R_k(l))(p^2+R_k(p))(p'^2+R_k(p'))}. \nonumber\\
\eea
Similarly as described above, the first two terms of (\ref{Eq:conn2phi}) do not contribute in (\ref{Eq:rhs}), but it turns out that the third one does. By performing the following series of variable changes in (\ref{Eq:conn2phi}): $q \rightarrow -q$, $q \rightarrow q-l$, $l \rightarrow l+q$ for the first term, and $q \rightarrow -q$, $q\rightarrow q-l+p$, $l\rightarrow l+p+q$ for the second term, we arrive at the following expression:
\bea
\label{Eq:rhsstrb}
\rhs&=&(-i)\frac{4e^3}{Z_{A,k}}\frac{Z^2_{e,k}}{Z_{\phi,k}}\int_p\theta (-p)\int_{q}\phi^{\dagger s}(q)\phi^s(q+p) \nonumber\\
&\times& \int_l R_k(l) \Big(f(l-p,l;l+q)-f(l,l+p;l+p+q)\Big). \nonumber\\
\eea
The $l$ integral has to be expanded in terms of $p$ and $q$ to let the combination $(p^2+2p\cdot q)$ emerge, in accordance with (\ref{Eq:leftfinal}). After a straightforward calculation, we arrive at
\bea
\label{Eq:rightfinal}
\rhs&=&(-i)\frac{4e^3}{Z_{A,k}}\frac{Z^2_{e,k}}{Z_{\phi,k}}\int_p\theta (-p)\int_{q}\phi^{\dagger s}(q)\phi^s(q+p) \nonumber\\
&\times& \frac{\xi_k}{4k^6} \big(p^2+2p\cdot q+{\cal O}(q^2,p^2q,p^3)\big)\int_l R_k(l).
\eea
Since $\int_l R_k(l)=2k^{d+2}\Omega_d/d(d+2)$, matching (\ref{Eq:rightfinal}) with (\ref{Eq:leftfinal}) we get
\bea
\label{Eq:mWTIfin}
Z_{e,k}/Z_{\phi,k}-1=\frac{Z_{e,k}^2}{Z_{\phi,k}^2}\frac{2e_k^2}{d(d+2)}\xi_k\Omega_d k^{d-4}.
\eea
The mWTI shows that in fact $Z_{e,k}\neq Z_{\phi,k}$, except for $\xi_k\equiv 0$ \cite{litim98}. This can only be satisfied at $d=4$, as we already had a constraint $\xi_k=2/(4-d)$ for $d\neq 4$. Note that, (\ref{Eq:mWTIfin}) is not due to any discrepancy between the flow equation and the mWTI, since by applying $k\partial_k$ to (\ref{Eq:mWTIfin}), we get
\bea
\label{Eq:Zratioflow}
\!\!\!k\partial_k \bigg(\frac{Z_{e,k}}{Z_{\phi,k}}\bigg)=\frac{Z_{e,k}^2}{Z_{\phi,k}^2}\frac{2e_k^2(d-4)}{d(d+2)}\xi_k\Omega_d k^{d-4}+{\cal O}(e^4),
\eea
which is in accordance with the flow equation (see Appendix C). 

We wish to emphasize that one cannot give up the constraint $\xi_k=2/(4-d)$ in favor of $\xi_k=0$, as the former is required by inner consistency between the flow equation and the ansatz of $\Gamma_k$ itself, and it is also a necessary condition to build compatibility between the gauge identity and the flow equation. Concerning the violation of the $Z_{e,k}=Z_{\phi,k}$ identity, there is no discrepancy between the mWTI and the flow equation; they lead to compatible results. We thus conclude that a possible way to circumvent the problem is what we did in Sec. II. in the first place, i.e., to project the flow equation onto a subspace of coupling flows, where the identity $Z_{e,k}=Z_{\phi,k}$ is imposed by hand.

\section{Conclusions}

In this paper we investigated the renormalization group flows of the couplings of the Abelian Higgs model with an $N$-component complex scalar field. We showed that a careful choice of the IR regularization scheme is necessary to find such an approximate solution of the flow equation that displays two charged fixed points for arbitrary $N$ in $d=3$ dimensions. We found that one of them is IR stable and thus capable of describing a second order phase transition, in particular, the superconducting transition for $N=1$. The other, being a tricritical fixed point controls the region of the parameter space that is attracted by the former one. The applied regulator has the property that it modifies eigenvalues of the propagators only in the respective Gaussian parts, and completely leaves the remaining momentum dependence untouched. We argued that this choice leads to better convergence properties of the effective action in terms of the derivative expansion.

Gauge symmetry and the corresponding Ward-Takahashi identities were also analyzed in detail. First, we found that in order to accommodate the flow equation and the applied ansatz of the effective action, except for $d=4$, one has to fix the gauge fixing parameter as $\xi_k \equiv 2/(4-d)$. The necessity of such a choice could be also explained via the violation of the Ward-Takahashi identity of the longitudinal photon due to IR regulator terms. We have also found violation of the $Z_{e,k}=Z_{\phi,k}$ (scalar) identity, which is only cured by a choice of $\xi_k \equiv 0$, if $d\neq 4$. This leads to the conclusion that the gauge and scalar Ward-Takahashi identities cannot be satisfied at the same time (except for $d=4$). Since the $\xi_k=2/(4-d)$ gauge is necessary from the point of view of the inner consistency between the flow equation and the applied ansatz (and similarly, compatibility between the flow equation and the mWTI), a possible strategy is to solve the flow equation in a subspace where $Z_{e,k}=Z_{\phi,k}$ is imposed by hand.

Our findings also showed that the flow equation and the modified Ward-Takahashi identities are compatible with each other. It is important to stress that the latter contains more information on the couplings, as one is free to extract the boundary conditions at the UV scale $\Lambda$ from them.

It would be interesting to improve the current truncation of the effective action, and include higher order terms in the derivative expansion. One would also need to optimize the flow and find the appropriate choice of regulator for those improved truncations. The current method, conveniently skipping any reference to background gauge fields might also be applicable to other scalar theories with $U(1)$ gauge symmetry, and one might also be interested in the case of non-Abelian gauge groups, in particular in investigating anomalous violation of the WTIs in general dimension $d$.

\section*{Acknowledgements}

T. H. was partially supported by RIKEN iTHES Project, the iTHEMS Program, and JSPS Grants No. 25287066 and No. 15H03663.

\makeatletter
\@addtoreset{equation}{section}
\makeatother 

\renewcommand{\theequation}{A\arabic{equation}} 

\appendix
\section{Calculation of loop integrals}

In this appendix we calculate two loop integrals. The first one appeared in calculating the flow of the effective action in a classical field $A_i$; see (\ref{Eq:gaugeflowb}). The other one is related to the gauge mWTI; see (\ref{Eq:rhsfinal}). 

Let us begin with (note that $m_k^2$ is set to zero)
\bea
&&I_1(p)=I_1(0)+\Delta I_1(p)=\nonumber\\
&&\tilde{\partial}_k\int_q \bigg[ \frac{2\delta_{ij}}{q^2+R_k(q)}-\frac{(p+2q)_i(p+2q)_j}{[q^2+R_k(q)][(q+p)^2+R_k(q+p)]}\bigg]. \nonumber\\
\eea
First we calculate $I_1(0)$.
\bea
I_1(0)&=&\tilde{\partial}_k\int_q \bigg[ \frac{2\delta_{ij}}{q^2+R_k(q)}-\frac{4q_iq_j}{[q^2+R_k(q)]^2}\bigg] \nonumber\\
&=&-\int_{|q|<k} \frac{4k^2\delta_{ij}-16q_iq_j}{k^5}\nonumber\\
&=&-\frac{\delta_{ij}}{k^5}\int_{|q|<k}(4k^2-16q^2/d),
\eea
where we used that $\int_q f(q^2) q_iq_j = \int_q f(q^2) q^2\delta_{ij}/d$. Then,
\bea
I_1(0)=-\frac{d-2}{d(d+2)}\frac{4}{k^{3-d}}\Omega_d \delta_{ij},
\eea
where $\Omega_d=2/[(4\pi)^{d/2}\Gamma(d/2)]$. The piece with momentum dependence is
\bea
\label{Eq:deltaI1}
&&\Delta I_1(p)=\nonumber\\
&&\int_{|q|<k}\bigg[\frac{4k(p+2q)_i(p+2q)_j}{k^4[p^2+2|p||q|x+q^2+R_k(p+q)]}-\frac{16kq_iq_j}{k^6}\bigg], \nonumber\\
\eea
where $x=\cos\theta$, $\theta$ being the angle between $p_i$ and $q_i$. We split the first term into two parts, depending on the regulator being zero or nonzero. Since $R_k$ is defined as $R_k(q)=(k^2-q^2)\Theta(k^2-q^2)$, $R_k(p+q)$ is nonzero only if
\bea
\label{Eq:constr}
q^2-2|p||q|x+p^2-k^2<0
\eea
is satisfied. That is, if $|q|<q_+\equiv k-|p|x+\frac{x^2-1}{2k}p^2+{\cal O}(p^3)$. Note that, since we are looking for an expansion in $p$ (i.e., $p$ can be considered infinitesimal), if $x<0$, then (\ref{Eq:constr}) is always satisfied (because $|q|<k$). Therefore, (\ref{Eq:deltaI1}) can be written as
\bea
\Delta I_1(p)&=&\frac{4}{k^5} \int_{|q|<k}(p+2q)_i(p+2q)_j\Theta(x<0) \nonumber\\
&+&\frac{4}{k^5}\int_{|q|<q_+}(p+2q)_i(p+2q)_j \Theta(x>0) \nonumber\\
&+&\frac{4}{k^3}\int_{q_+<|q|<k} \frac{(p+2q)_i(p+2q)_j}{p^2+2|p||q|x+q^2}\Theta(x>0) \nonumber\\
&-&\frac{16}{k^5}\int_{|q|<k} q_iq_j.
\eea
This can be reformulated as
\bea
\label{Eq:deltaI1b}
\Delta I_1(p)&=&\frac{4}{k^5}\int_{|q|<k} (p+2q)_i(p+2q)_j \nonumber\\
&+&\frac{4}{k^3}\int_{q_+<|q|<k} (p+2q)_i(p+2q)_j \Theta(x>0) \nonumber\\
&\times& \bigg[\frac{1}{p^2+2|p||q|x+q^2}-\frac{1}{k^2}\bigg]-\frac{16}{k^5}\int_{|q|<k} q_iq_j. \nonumber\\
\eea
Performing the radial integral, and expanding all terms up to ${\cal O}(p^2)$, (\ref{Eq:deltaI1b}) leads to
\bea
\label{Eq:deltaI1c}
\Delta I_1(p)&=&\frac{4k^{d-5}}{d}p_ip_j \int_\Omega 1 -16k^{d-5}p^2 \int_\Omega x^2 \hat{q}_i\hat{q}_j \Theta(x>0), \nonumber\\
\eea
where $\int_\Omega = \int d\Omega (2\pi)^{-d}$, and $\hat{q}_i=q_i/|q|$. Since $x=\cos \theta$ can be written as $x=\hat{q}_i\hat{p}_i$, the second term in (\ref{Eq:deltaI1c}) needs the evaluation of the integral
\bea
\int_\Omega \hat{q}_i \hat{q}_j \hat{q}_k \hat{q}_l \Theta(x>0).
\eea
The $\Theta$ function simply restricts the surface integral to a half-(unit)sphere. Because of symmetry, under the integral it can be substituted as $\Theta(x>0)\rightarrow 1/2$. One then makes use of the identity
\bea
\label{Eq:deltaid}
\!\!\!\!\int_\Omega \hat{q}_i \hat{q}_j \hat{q}_k \hat{q}_l = \frac{\Omega_d }{d(d+2)}(\delta_{ij}\delta_{kl}+\delta_{ik}\delta_{jl}+\delta_{il}\delta_{jk}),
\eea
and arrives at
\bea
\Delta I_1(p)=\frac{8\Omega_dk^{d-5}}{d(d+2)}\Big(\frac{d-2}{2}p_ip_j-p^2\delta_{ij}\Big).
\eea
This results in
\bea
\!\!\!\!I_1 (p)&=&-\frac{4\Omega_dk^{d-3}}{d(d+2)}(d-2) \delta_{ij}\nonumber\\
&-&\frac{8\Omega_dk^{d-5}}{d(d+2)}\Big(p^2\delta_{ij}-\frac{d-2}{2}p_ip_j\Big)+{\cal O}(p^4),
\eea
which leads to (\ref{Eq:gaugeflow2}).

The second integral we are performing in this appendix is
\bea
I_{2,i}(p)&=&\int_q \bigg[ \frac{(2q+p)_i}{[q^2+R_k(q)][(q+p)^2+R_k(q+p)]}\nonumber\\
&-&\frac{(2q-p)_i}{[q^2+R_k(q)][(q-p)^2+R_k(q-p)]}\bigg]R_k(q),\nonumber\\
\eea
which arises after plugging (\ref{Eq:conn}) into (\ref{Eq:rhs}) at $m^2_k=0$. Since $R_k(q)=R_k(-q)$, after performing a $q\rightarrow -q$ variable change in the second term, we get
\bea
I_{2,i}(p)=\int_{q} \frac{2(2q+p)_iR_k(q)}{[q^2+R_k(q)][(q+p)^2+R_k(q+p)]}. \nonumber\\
\eea
First, we exploit that $R_k(q)/[q^2+R_k(q)]=1-q^2/k^2$, if $|q|<k$, and then split the integral similarly as above:
\bea
I_{2,i}(p)&=&\frac{2}{k^2}\int_{|q|<k}(2q+p)_i (1-q^2/k^2)\nonumber\\
&+&2\int_{q_+<|q|<k} (2q+p)_i (1-q^2/k^2)\Theta(x>0) \nonumber\\
&\times&\bigg[\frac{1}{q^2+p^2+2|q||p|x}-\frac{1}{k^2}\bigg].
\eea
Performing the radial integral, and then expanding everything up to ${\cal O}(p^3)$ one arrives at
\bea
I_{2,i}(p)&=&\frac{4k^{d-2}}{d(d+2)}p_i\int_\Omega 1-\frac{8k^{d-4}}{3}p^3\int_\Omega \hat{q}_ix^3\Theta(x>0) \nonumber\\
&+&{\cal O}(p^5).
\eea
Under the integral we again substitute $\Theta(x>0)\rightarrow 1/2$, and after making use of (\ref{Eq:deltaid}), we get
\bea
\label{Eq:I2final}
\!\!I_{2,i}(p)&=&\frac{4\Omega_dk^{d-2}}{d(d+2)}p_i-\frac{4\Omega_dk^{d-4}}{d(d+2)}p^2p_i+{\cal O}(p^5).
\eea
This completes the derivation as (\ref{Eq:I2final}) leads to (\ref{Eq:rhsfinal}). 

\renewcommand{\theequation}{B\arabic{equation}} 

\section{Calculations with different regulators}

In this part of the Appendix we evaluate the two forms of the flow equation related to regulator choices R1 and R2 [i.e., (\ref{Eq:flowlog2}) and (\ref{Eq:flowlogb})]. We introduce the following notations for the eigenvalues (\ref{Eq:eigen2}):
\bea
\label{Eq:eigen3}
\!\!\!\!\!\!\gamma_{k,2}^{(i)}(q)=Z_k^{(i)}q^2+\frac{A_k^{(i)}}{q^2}(\partial_j\sigma)^2+\frac{B_k^{(i)}}{q^2}(\partial_j\sigma)^2\cos^2\theta,
\eea
where the $k$-dependent constants $Z_k^{(i)}$, $A_k^{(i)}$, and $B_k^{(i)}$ can be read off from the respective expressions. Let us analyze ({\ref{Eq:flowlogb}) first. Plugging (\ref{Eq:eigen3}) into (\ref{Eq:flowlogb}), after performing the radial integral, we get
\bea
\label{Eq:R1}
\!\!\!\!\partial_k &&\!\!\!\!\!\!\!\!\Gamma_k^{\Rone}|_{\partial_j \sigma}=\int_x\int_\Omega\sum_{i=1,2,d+2N} \sum_{n=0}^{\infty}\frac{(-1)^nk^{d-1-4n}}{(d-2n)(Z_k^{(i)})^n}\nonumber\\
&\times&(A_k^{(i)}+B_k^{(i)}\cos^2\theta)^n (\partial_j \sigma)^{2n} \nonumber\\
&-&\int_\Omega \frac{\pi k^{-d-1}}{2\sin(d\pi/2)} \left(\frac{A_k^{(i)}+B_k^{(i)}\cos^2\theta}{Z_k^{(i)}}\right)^{d/2}(\partial_j \sigma)^{d}.\nonumber\\
\eea
The angular integrals can be performed analytically, but for compactness we left them undone. First, one notes that if $d$ is even, in the sum of the right-hand side the term for which $n=d/2$ is divergent. Note that, since if $d\rightarrow 2n$, $\sin(d\pi/2)\approx (-1)^n\pi(d-2n)/2$, this is always canceled by the last term; thus, the expression is well defined. If $d$ is odd, then the sum is finite, but we are left with the second term being non analytic in $\partial_j \sigma$. Without proof, we expect that these type of terms vanish, once one goes to higher orders in $\partial_j \sigma$ in the eigenvalues (\ref{Eq:eigen3}). This expectation is based on the analyticity of the effective action, and on the fact that the result of integrals 
\bea
\int_0^k dq q^{d-1} \big(k^2+\const\times[(\partial_i\sigma)^2/q^2]^m\big)^{-1}
\eea
contains a term proportional to $(\partial_i\sigma)^{d}$ for all $m$ [all remaining contributions are of ${\cal O}\big((\partial_i \sigma)^{2j}\big), j\geq m$]. That is to say, for calculating the term ${\cal O}\big((\partial_j \sigma)^d\big)$ in (\ref{Eq:R1}), one should go to all orders in (\ref{Eq:eigen3}) in terms of $\partial_j \sigma$. Since $\Gamma_k$ is an analytic function of $\partial_j \sigma$, these terms should ultimately add up to zero when $d$ is odd.

(\ref{Eq:R1}) can be summarized as
\bea
\label{Eq:R1b}
\!\!\!\!\partial_k \Gamma_k^{\Rone}|_{\partial_j \sigma}&=&\int_x\int_\Omega\sum_{i=1,2,d+2N} \sum_{n\neq d/2}\frac{(-1)^nk^{d-1-4n}}{(d-2n)(Z_k^{(i)})^n}\nonumber\\
&\times&(A_k^{(i)}+B_k^{(i)}\cos^2\theta)^n (\partial_j \sigma)^{2n}.
\eea
On the other hand, for (\ref{Eq:flowlog2}) one arrives at
\bea
\label{Eq:R2b}
\!\!\!\!\partial_k \Gamma_k^{\Rtwo}|_{\partial_j \sigma}&=&\int_x\int_\Omega\sum_{i=1,2,d+2N} \sum_{n}\frac{(-1)^n2k^{d-1-4n}}{d(Z_k^{(i)})^n}\nonumber\\
&\times&(A_k^{(i)}+B_k^{(i)}\cos^2\theta)^n (\partial_j \sigma)^{2n}.
\eea
The obtained results are very much alike, but nevertheless different. For the prediction of the wave function renormalization, we have to take the $n=1$ term in each sum, which leads to (\ref{Eq:waveR1}) and (\ref{Eq:waveR2}).

\renewcommand{\theequation}{C\arabic{equation}} 
\vspace{1cm}
\section{Flow of $Z_{e,k}$}

Throughout the paper we assumed that $Z_{e,k}=Z_{\phi,k}$, but the mWTI in Sec. IV revealed that it can only be maintained at $\xi_k=0$. Here we show that the same result can be obtained directly from the flow equation. In order to calculate $k\partial_k Z_{e,k}$, one has to project the flow equation (\ref{Eq:floweq2}) onto the operator $\sim \sigma^a \pi^aA_i$. The appropriate projection of the lhs of (\ref{Eq:floweq2}) is
\bea
k\partial_k \Gamma_k|_{\sigma^a\pi^aA_i}(p,-p)=2iep_i k\partial_k Z_{e,k},
\eea
where we set the gauge momentum to zero for simplicity. The corresponding terms in the rhs are
\begin{widetext}
\bea
\label{Eq:C2}
(-i)\frac{Z_{e,k}^3e^3}{Z_{A,k}Z_{\phi,k}^2}\tilde{\partial}_k\bigg[\int _l \frac{2(2p-l)^2(l-p)_i+2(1-\xi_k)(2p\cdot l-l^2)^2(l-p)_i/l^2}{\big(l^2+R_k(l)\big)\big((p-l)^2+R_k(p-l)\big)^2}+\frac{(2p-l)_i-(1-\xi_k)(2p\cdot l -l^2)l_i/l^2}{\big(l^2+R_k(l)\big)\big((p-l)^2+R_k(p-l)\big)}\bigg], \nonumber\\
\eea
\end{widetext}
which has to be expanded up to ${\cal O}(p)$ to arrive at
\bea
\label{Eq:C3}
\frac{k\partial_k Z_{e,k}}{Z_{e,k}}&&=\nonumber\\
&&\!\!\!\!\!\!\!\!\!\!\!\!\!e_k^2\frac{Z_{e,k}^2}{Z_{\phi,k}^2}\bigg[\frac{16(d-1)}{d^2}-\frac{2}{d^2}\frac{3d^2+4d-16}{d+2}\xi_k\bigg]\Omega_d k^{d-4}. \nonumber\\
\eea
The flow of $Z_{\phi,k}$ slightly changes for the choice $\tilde{\sigma}^a=0$, and an analogous calculation as of Sec. III leads to
\bea
\label{Eq:C4}
\frac{k\partial_k Z_{\phi,k}}{Z_{\phi,k}}=e_k^2\frac{Z^2_{e,k}}{Z^2_{\phi,k}}\bigg[\frac{16(d-1)}{d^2}-\frac{2}{d^2}\frac{4d^2-16}{d+2}\xi_k\bigg] \Omega_d k^{d-4}.\nonumber\\
\eea
Note that, (\ref{Eq:C4}) agrees with (\ref{Eq:C3}) for $d=4$, but otherwise one gets
\bea
\label{Eq:flowward}
k\partial_k \bigg(\frac{Z_{e,k}}{Z_{\phi,k}}\bigg)=\frac{Z_{e,k}^3}{Z_{\phi,k}^3}\frac{2e_k^2(d-4)}{d(d+2)}\xi_k \Omega_d k^{d-4},
\eea
which is compatible with the mWTI [see (\ref{Eq:Zratioflow})], but note that, a multiplicative factor of $Z_{e,k}/Z_{\phi,k}$ appeared in the rhs of (\ref{Eq:flowward}) compared to (\ref{Eq:Zratioflow}). A similar factor appears when one confronts the flow equation and the mWTI regarding the gauge identity. Nevertheless, the arguments presented in this appendix also shows that the only choice to maintain the scalar identity (i.e. $Z_{e,k}=Z_{\phi,k}$) is to take $\xi_k=0$.

\end{document}